\documentclass[11pt]{article}

\usepackage[left=2cm, right=2cm, top=2.5cm, bottom=2.5cm]{geometry}
\usepackage[x11names]{xcolor}
\usepackage{fancyhdr, amssymb, cancel, amsmath, graphicx, pgfplots, tikz}

\newcommand{\stylecolor}{black}

\usepackage[colorlinks=true, urlcolor=orange, linkcolor=blue, citecolor=red, hyperindex=true, linktocpage=true]{hyperref}

\pgfplotsset{samples=200}

\usepackage[explicit]{titlesec}

\newcommand*\sectionlabel{}
\titleformat{\section}
  {\gdef\sectionlabel{}
   \Large\bfseries\scshape}
  {\gdef\sectionlabel{\thesection. }}{0pt}
  {\begin{tikzpicture}[remember picture,overlay]
	\draw (-0.2, 0) node[right] {\textsf{\sectionlabel#1}};
	\draw[thick] (0, -0.4) -- (\textwidth, -0.4);
       \end{tikzpicture}
  }
\titlespacing*{\section}{0pt}{15pt}{20pt}

\newcommand*\subsectionlabel{}
\titleformat{\subsection}
  {\gdef\subsectionlabel{}
   \large\bfseries\scshape}
  {\gdef\subsectionlabel{\thesubsection.\ \  }}{0pt}
  {\begin{tikzpicture}[remember picture,overlay]
    	\draw (-0.15, 0) node[right] {\textsf{\subsectionlabel#1}};
       \end{tikzpicture}
  }
\titlespacing*{\subsection}{0pt}{10pt}{10pt}

\pgfplotsset{every axis legend/.append style={at={(1.02,1)},anchor=north west}}

\newcommand{\titletext}{Binary decision making with very heterogeneous influence}

\begin{document}
\thispagestyle{empty}

\begin{equation*}
\begin{tikzpicture}
\draw (0.5\textwidth, -3) node[text width = \textwidth] {{\huge \begin{center} \color{\stylecolor} \textsf{\textbf{\titletext}} \end{center}}}; 
\end{tikzpicture}
\end{equation*}
\begin{equation*}
\begin{tikzpicture}
\draw (0.5\textwidth, 0.1) node[text width=\textwidth] {\large \color{black} $\text{\textsf{Andrew Lucas}}$};
\draw (0.5\textwidth, -0.5) node[text width=\textwidth] {\small  \textsf{Department of Physics, Harvard University, Cambridge, MA 02138, USA}};
\end{tikzpicture}
\end{equation*}
\begin{equation*}
\begin{tikzpicture}
\draw (0.5\textwidth, -6) node[below, text width=0.8\textwidth] {\small  We consider an extension of a binary decision model in which nodes make decisions based on influence-biased averages of their neighbors' states,  similar to Ising spin glasses with on-site random fields.   In the limit where these influences become very heavy-tailed, the behavior of the model dramatically changes.   On complete graphs, or graphs where nodes with large influence have large degree, this model is characterized by a new ``phase" with an unpredictable number of macroscopic shocks, with no associated critical phenomena.   On random graphs where the degree of the most influential nodes is small compared to population size, a predictable ``glassy" phase without phase transitions emerges.  Analytic results about both of these new phases are obtainable in limiting cases.   We use numerical simulations to explore the model for more general scenarios.  The phases associated with very influential decision makers are easily  distinguishable experimentally from a homogeneous influence phase in many circumstances, in the context of our simple model.};  
\end{tikzpicture}
\end{equation*}
\begin{equation*}
\begin{tikzpicture}
\draw (0, -13.1) node[right] {\texttt{lucas@fas.harvard.edu}};
\draw (\textwidth, -13.1) node[left] {\textsf{\today}};
\end{tikzpicture}
\end{equation*}

\tableofcontents

\pagestyle{fancy}
\renewcommand{\headrulewidth}{0pt}
\fancyhead{}

\fancyhead[L] {\textsf{\titletext}}
\fancyhead[R] {\textsf{\thepage}}
\fancyfoot{}

\section{Introduction}
Qualitative understanding of the behavior of many social systems has been afforded by statistical physics and the study of systems on complex networks \cite{Barrat2008, Castellano2009, Samanidou2007}.   In particular, one of the major hopes of sociophysics and econophysics is to observe a ``thermodynamic"  limit in markets and social dynamics as the population size gets large \cite{Bouchaud2008}, similarly to how a thermodynamic limit arises in a typical disordered system such as a spin glass \cite{DeDominicis2006}, which may allow for quantitative predictions about the behavior of social systems.    One of the most interesting applications of statistical physics in this field is to the study of market crashes and other phase transitions in decision making processes.  These are phenomena which are challenging to describe using classical economics, yet easily understood -- at least conceptually -- from the perspective of disordered systems \cite{Bouchaud2008}.

Recently, \cite{Lucas2013} introduced a binary decision model which admitted a very simple solution on typical random graphs.  The decision model proposed was a variant of the random field Ising model which has been used frequently to model decision making processes \cite{Bouchaud2013, Galam1991, Galam1997} due to disorder induced hysteresis and avalanche dynamics \cite{Sethna1993}.   Other models of cascades and global avalanches \cite{Crucitti2004, Watts2002} such as the fiber bundle model \cite{Pradhan2010} have also been employed, and share some qualitative features as well.   Although many of these models have focused on the avalanche dynamics near or at a phase transition, \cite{Lucas2013} emphasized in addition a seemingly universal glassy phase to these models, which was remarkably simple to understand on locally tree-like graphs.

The aim of this paper is to describe a modification of the model in \cite{Lucas2013} to allow for influential decision making, where a given node does not weight its neighbors equally when determining its own state.    Similar models using the random field Ising model were studied in the past \cite{Kacperski1999}, although not beyond the level of mean field theory.    Influential decision making has long been assumed in consumer research, for example: most prevalent is the so-called ``two-step flow" model, which assumes that information first passes through influential decision makers, who then filter it to a cluster of followers \cite{Coleman1957, Katz1955, Roch2005}.   However, the presence of influential decision makers is not required for shocks and large cascades, as pointed out by the physics community \cite{Watts2007}.   However, since the question of just how important influential individuals are in decision making processes is still open (for some recent experimental studies, see \cite{Bakshy2011, Cha2010, Zhou2011}), it is worth exploring more carefully the role that influence could play in decision making processes.  

In this paper, we provide a conceptual framework for understanding the effects of influence in decision making.   We show that interplay between network structure and the influence of given nodes allows for a variety of possible outcomes -- three of which exist as solvable limits, and which we detail in this paper.   The first possible outcome is that influence is a negligible effect, which gives us the phases of \cite{Lucas2013}.   The second possible outcome is unpredictable decision making, in which shocks occur without critical phenomena, and the decision making process is entirely controlled by a few nodes.   This limit occurs when there are nodes which are both heavily connected and have a large fraction of the global influence.    The third limit corresponds to the existence of a single, very ``glassy" phase, independent of the network structure, which corresponds to non-interacting clusters following single influential nodes.   This limit occurs when there are nodes that carry heavy influence but do not have large degree.

The outline of this paper is as follows.   In Sec. \ref{sec2} we describe our model, and in Sec. \ref{sec3} we review basic information about heavy-tailed distributions of random variables.   Sec. \ref{sec4} describes the unpredictable decision making phase in the simple case where the network is a complete graph, which dramatically simplifies the analysis.   Sec. \ref{sec5} describes the rich variety of scenarios which are possible on random graphs, including the limit when influence becomes weak.   We conclude by discussing the relevance of our results to possible experiments.

\section{Binary Decision Making with Influential Nodes} \label{sec2}
Let us recall the binary decision model presented in \cite{Lucas2013}, which is similar to the random field Ising model.   For more details and figures one can refer to this reference;  here we will discuss the highlights and not include derivations of any results. 

We consider an undirected graph with nodes labeled by $v$.  We will denote the degree of $v$, which is the number of edges entering $v$, with $k_v$.    On each vertex, we place a bit $x_v \in \lbrace 0,1\rbrace$ for each vertex $v$.    The phrase binary decision making comes from the fact that there only are two possible states, and no intermediate ``wavering" state.   The rules are that \begin{equation}
x_v = \Theta(P_v h(Q_v)-p) \label{eq1}
\end{equation}where $\Theta(x)$ is the step function which is 1 if $x\ge 0$ and 0 if $x<0$, $p$ is a global ``magnetic field" and $P_v$ correspond to quenched random variables (the ``internal magnetic fields").   $h$ is a monotonically increasing function which we take to be identical for all nodes.    In this paper, we always take \begin{equation}
h(Q) = 1+AQ
\end{equation}for $A>0$, for simplicity.   Finally, $Q_v$ corresponds to the fraction of the neighbors $u$ of the node $v$ which are in state 1: \begin{equation}
Q_v = \mathrm{P}(x_u=1|u, v \; \mathrm{neighbors}).   \label{Qvold}
\end{equation}The intuition for Eq. (\ref{eq1}) is simple: each node compares the global field $p$ to some internal field $P_v h(Q_v)$:  social interactions imply that the internal field it sees is dependent on the actions of its neighbors.   If the global field is smaller than this internal field, it will adopt state 1; otherwise it will adopt state 0.   As a simple example, we can imagine modeling someone choosing to buy a good ($x_v=1$) if the price $p$ is below the perceived value of that good $P_v h(Q_v)$, and not buying ($x_v=0$) otherwise.  This model has non-trivial behavior because not only is $P_v$ different for each node, but the perceived values are dependent on what the states of other nodes.

On a complete graph, or in mean-field approximation, the solution to this model is quite simple:  it is straightforward to see that if \begin{equation}
F(P) = \mathrm{P}(P_v > P),
\end{equation}that the mean field states, characterized by an overall fraction $q$ of states in state 1, are determined by the solutions to the equation \begin{equation}
q = F\left(\frac{p}{h(q)}\right).   \label{mfeq}
\end{equation}
Numerically, Eq. (\ref{eq1}) can be solved from an arbitrary initial condition by simply iterating through every single node, and flipping its state whenever Eq. (\ref{eq1}) is violated.   This will always converge to an equilibrium state; though as we will discuss shortly, there are generically many equilibria that can be reached.   A drastic simplification occurs whenever we can \emph{start} from an equilibrium state:  in this case, if we increase $p$, we know that the only nodes which will flip are nodes that will flip from 1 to 0, and oppositely if we decrease $p$.    In this paper our simulations will exploit this simplification.

It is often the case that Eq. (\ref{mfeq}) admits multiple solutions.   A typical scenario (referred to henceforth as bistable) is that there are three solutions:  $q_1(p)<q_2(p)<q_3(p)$, with $q_1$ and $q_3$ stable equilibria, and $q_2$ an unstable equilibrium.   These solutions certainly depend on the value $p$, as the notation suggests.    This bistability is analogous to the way that the simple Ising model in an external magnetic field can admit multiple stable equilibria.    Furthermore, as we increase the value of $p$, we usually find that eventually, the solutions $q_2$ and $q_3$ approach each other, and simultaneously disappear.    If one was sitting at the stable equilibrium $q_3$ as $p$ increased,  past a critical point one would suddenly find that $q_1<q_3$ is the only stable equilibrium -- this corresponds to a discontinuous phase transition, or in economic language, a market crash.    This is analogous to the behavior of the standard Ising model, and is not surprising.   One can study the stability of solutions to Eq. (\ref{mfeq}) by studying the parameter \begin{equation}
\alpha \equiv F^\prime \left(\frac{p}{h(q)}\right) \frac{ph^\prime(q)}{h(q)^2}.
\end{equation}If $\alpha<1$, the equilibrium is stable, and if $\alpha>1$ it is unstable.   In fact, there is a microscopic understanding of this result:   one can show that the probability that flipping node $u$'s state will cause one of its neighbors to flip to the state is equal to $\alpha/\langle k\rangle$, where $\langle k\rangle$ is the expected degree (average number of edges) of a node in the graph.   From here, one finds under typical circumstances that the expected size of an avalanche -- the cumulative number of nodes who have flipped their state, due to the flip of a single node -- is $1/(1-\alpha)$, so long as $\alpha<1$.   Because this is a finite effect, the equilibrium is thus stable when $\alpha<1$.   In such a regime, the effect of such an avalanche can be modeled under linear response, which is simply a way of saying that all macroscopic perturbations to $q$ need only be kept track of to lowest order, so long as the number of nodes in the graph, which we call $N$, is large.

More exciting within the economic context is the emergence of complexity:  one finds that on any graph with finite $\langle k\rangle$, so long as $\alpha>0$, the number of stable equilibria is exponential in the size of the graph.    Due to the ``glassy" nature of the random field Ising model, this is not particularly strange, although we find especially elegant formulas describing this complexity in the context of this simple model.  The way to understand the emergence of complexity in this model is as follows: consider a pair of neighboring states which are both very close to the point where $P_vh(Q_v)=p$, and which can flip if any of their neighbors flips: this happens with probability $\alpha/\langle k\rangle$.   Since there are $\langle k\rangle$ edges per node, the probability that any one node can simultaneously flip with one of its neighbors is then roughly $\langle k\rangle (\alpha/\langle k\rangle)^2$, and there is a  stable equilibrium where the pair of nodes is 0, and an equilibrium where the pair is 1.    In particular, one can show that there is a \emph{macroscopic spectrum} of allowed values of $q$ associated to any single stable phase, at mean field level, whose width is \begin{equation}
\Delta q = \frac{\alpha^2}{(1-\alpha)\langle k\rangle}   \label{eq7}
\end{equation}
so long as $\langle k\rangle$ is not too small, by accounting for the possibility of avalanches due to node flips.    This relation follows more rigorously from Thouless-Anderson-Palmer (TAP) equations, and requires that $\langle k\rangle$ be large enough so that the linear response regime is valid.    A formal proof of whether or not this phase is truly a ``glass" is subtle, and we will comment more later.    

In this paper, we will provide a new twist on this model:  we will alter Eq. (\ref{Qvold}) to be a weighted average over the neighbors of $v$, in the following sense.    To each node, we assign an influence parameter $J_v$, and we let \begin{equation}
Q_v = \frac{ \sum_u J_u x_u}{ \sum_u J_u }   \label{Qvnew}
\end{equation}The sums in Eq. (\ref{Qvnew}) are over each node that is connected to node $v$.    It is straightforward to see that within mean field theory, where we replace all $x_v$ in the equations above by $\langle x_v\rangle \equiv q$, that we find that $Q=q$ (using Eq. (\ref{Qvnew})), which implies that the mean field equations are exactly the same as they are in the homogeneous influence case.    However, it is possible to find distributions on $J_v$ which make mean field theory a bad approximation, even on complete graphs.   We will show below that if the probability distribution on $J_v$ is scale free with a small enough exponent, then in fact a few nodes dominate $Q$, and the resulting model behaves dramatically different from mean field theory.    This will be the subject of much of this paper.

Before describing the results for the binary decision model under very heavy influence distributions, let us return to the question of whether or not this model is a true glass.   A mathematical approach to this problem is to look for divergent spin glass susceptibility.   Here we simply quote a theorem: for any model which admits an energy functional of the form \begin{equation}
H = - \frac{1}{2}\sum_{u\ne v} J_{uv} x_u x_v - \sum_v m_v x_v,  \label{hising}
\end{equation}one can rigorously show that so long as each $J_{uv}\ge 0$, in a thermal ensemble where the probability of drawing a given state in equilibrium is proportional to $\mathrm{e}^{-H/T}$, for some temperature $T$, that no spin glass phase exists \cite{Krzakala2010}.   In the special case of $h(Q)=1+AQ$ we consider in the numerical simulations of this paper, there does exist an energy functional with the property that its minimization corresponds to the solutions of Eq. (\ref{eq1}).   To see this, divide through Eq. (\ref{eq1}) by $P_v/J_{\mathrm{neighbors}}J_v$, where $J_{\mathrm{neighbors}}$ corresponds to the sum of the influence of each of the neighbors of $v$.   One finds that Eq. (\ref{eq1}) becomes \begin{equation}
x_v = \Theta\left(\sum A J_v J_u x_u + \left[1-\frac{p}{P_v}\right]J_v\sum_u J_u \right),
\end{equation}which can be seen easily to correspond exactly to minimizing the energy functional \begin{equation}
H = -\frac{1}{2}\sum_{u,v\;\mathrm{ neighbors}} AJ_uJ_v x_ux_v - \sum_v \left(1-\frac{p}{P_v}\right)J_vx_v\sum_{u,v\;\mathrm{neighbors}} J_u. \label{efunc}
\end{equation}
So by one definition, this model for $h(Q)=1+AQ$ does not describe a glass.    Note that for other choices of $h(Q)$, such an energy functional may not exist.

However, the definition of glass above is too restrictive in our context.  To see this, note that the previous paragraph referred to the Gibbs distribution at finite temperature $T>0$, implying that the noisy ``thermal" dynamics of the system admit a stationary Gibbs distribution.   This is a non-trivial and overly restrictive statement in the context of sociophysics.  In particular, at $T=0$ (the non-dynamical limit in which we are interested), we generically expect most of the stable equilibria described above will have vanishing weight in the Gibbs distribution since they are not necessarily states of minimal $H$.   However, from the equations of motion, which is how we originally defined the model, there is no intuitive reason why these equilibria must be suppressed; one can construct a Markov chain whose ``zero temperature" limit does not suppress so many stable equilibria.    One would therefore like a  definition of glasses, appropriate for sociophysics, which is more general: a heuristic but physically motivated definition would be a system with a complexity paradigm that admits dynamics on arbitrarily large time scales (perhaps breaking ergodicity).   Our model, in many phases, has a complex landscape of exponentially many solutions in many situations.   It is reasonable to suspect that  large distributions in the sizes of ``glassy clusters" whose flips provide the exponential numbers of equilibria lead to mixing dynamics over many time scales, when one studies the dynamics of this model.  For this reason, we will drop the quotes around the word ``glass" and refer to phases of this model with exponential numbers of equilibria as glassy phases.   One should note however that these states do not necessarily obey one conservative definition of a (spin) glass.

As a final note, for the remainder of this paper, we will make the approximation that the sums in Eq. (\ref{Qvnew}) are over the neighbors $u$ of the node $v$, as well as the node $v$ itself.   The addition of $v$ to this sum makes some of the analytic calculations of this paper simpler, but does not lead to any qualitatively new physics, as we have checked numerically.  

\section{L\'evy Flights} \label{sec3}
In this section, we will focus on influence statistics when the influence is distributed according to a power law distribution with a very heavy tail -- in this paper, the word ``very" will quantify distributions with infinite mean.   Since there are many experimental studies suggesting reasonable measures of social influence have power law tails \cite{Adamic2000, Clauset2009, Price1965, Ebel2002, Newman2004, Zhou2011}, this is not an unreasonable toy model.    In particular, in some circumstances \cite{Zhou2011} found very heavy tailed ``influence" distributions (out-degrees of nodes in a network) in social recommender systems in social media. 
  
The extreme value statistics of random variables with infinite mean, described in terms of  the well-known ``L\'evy flights", have found a variety of interesting applications in physics \cite{Bertin2005}, primarily as descriptions of statistical systems which cannot reach an equilibrium distribution \cite{Bardou2002, Bouchaud1992}.   We will review some of the basic calculations as we will wish to derive some less commonly known results as well.    Then we will describe how these L\'evy flights are manifested in our binary decision model as two alternative solvable phases of our binary decision model.    Some of the calculations below are described in an alternate fashion in \cite{Bardou2002} -- we repeat them here for simplicity.  
\subsection{Influence Statistics} 
Let us consider the following simple problem.   Suppose we have $N$ independently and identically distributed (iid) random variables $J^i$  ($i=1,\ldots,N$) distributed according to the distribution \begin{equation}
\mathrm{P}(J^i>x) = \max(1,x)^{-\nu}   \label{jdist}
\end{equation}for $0<\nu < 1$.   For the remainder of this section, we will assume that $x\ge 1$, to simplify notation.   Without loss of generality, we can re-label these variables as $J_i$, with the re-labeling chosen so that $J_1>J_2>\cdots > J_N$.\footnote{With probability 1, since this is a continuous distribution, all $J_i$ will be distinct.}   The cumulative distribution function for $J_k$ is easy to find: \begin{align}
\mathrm{P}(J_k<x) &= \left(\begin{array}{c} N \\ k-1 \end{array}\right) \left(x^{-\nu}\right)^{k-1}\left(1-x^{-\nu}\right)^{N-k+1} + \left(\begin{array}{c} N \\ k-2 \end{array}\right) \left(x^{-\nu}\right)^{k-2}\left(1-x^{-\nu}\right)^{N-k+2}+\cdots  \notag \\
&= \left(1-x^{-\nu}\right)^N \sum_{j=0}^{k-1}  \left(\begin{array}{c} N \\ j \end{array}\right) \left(\frac{1}{x^\nu-1}\right)^j
\end{align}The interpretation is simple:  the first term corresponds to the probability that $J_{k-1}$ are all bigger than $x$ (the multiplicative factor corresponds to the number of possible permutations of $J_{k-1}$ from the original $J^\alpha$).   However, there is also the probability that $J_{k-1}$ is also smaller than $x$, etc., and accounting for all of these possibilities we obtain the total sum.    When $N$ is large and $x\gg 1$, we can make the approximation \begin{equation}
\mathrm{P}(J_k<x)  \approx \exp\left[-\frac{N}{x^\nu}\right] \sum_{j=0}^{k-1}   \left(\begin{array}{c} N \\ j \end{array}\right)  x^{-j\nu} \label{eq3}
\end{equation}
From Eq. (\ref{eq3}) we see that the scale for $J$ to be large is $N^{1/\nu}$:  in particular, the probability distribution on $J_1$ is rapidly suppressed (as an exponential in $N$) for $x\ll N^{1/\nu}$.   The probability that $J_1<N^{1/\nu}$ is $1-1/\mathrm{e}$.    The probability that there are exactly $k$ $J^i$s which are larger than $N^{1/\nu}$ is given by \begin{equation}
\mathrm{P}(k\text{ ``large" }J\mathrm{s}) = \mathrm{P}(J_{k+1}<N^{1/\nu})-\mathrm{P}(J_k<N^{1/\nu}) \approx \frac{1}{\mathrm{e}k!}
\end{equation}It is straightforward to argue that the scale of the typical $J_i<N^{1/\nu}$ is  \begin{equation}
\left\langle \sum_{i=1}^N J^i\Theta\left(N^{1/\nu}-J^i\right) \right\rangle \approx N \int\limits_1^{N^{1/\nu}} \mathrm{d}x \frac{\nu x}{x^{\nu +1}} \approx \frac{\nu}{1-\nu} N^{1/\nu}  \label{avgsmall}
\end{equation}with standard deviation \begin{equation}
\left[\mathrm{Var}\left( \sum_{i=1}^N J^i\Theta\left(N^{1/\nu}-J^i\right) \right)\right]^{1/2} \approx  \sqrt{\frac{\nu}{2-\nu}} N^{1/\nu}.   \label{stdsmall}
\end{equation}

These manipulations suggest that only a few influence variables dominate the sum $J_1+\cdots +J_N$.   It therefore seems reasonable to focus only on the largest few random variables in this sum.    In particular, we will focus on the random variables \begin{equation}
K_i  = \frac{J_i}{J_1+\cdots + J_N}
\end{equation}in this subsection, as these are the weights enter into Eq. (\ref{Qvnew}) for $Q$ in the binary decision model on complete graphs.  Of course, it really is no different on any other graph either, as although the denominator only contains $k_v+1$ terms for node $v$, the sum will still be dominated by only a few terms.    More importantly, while $J_i$ have divergent and badly-behaved distributions, the distributions on $K_i$ are well-behaved.   In general, it is very difficult to compute the distributions of $K_i$.   However, we notice that if $\nu$ is close to 0, Eqs. (\ref{avgsmall}) and (\ref{stdsmall}) suggest that all of the small $J$s added together are still significantly smaller than the largest few $J$s.    

Let's begin by looking at the distribution on the ratios $J_k/J_1$.    The cumulative joint distribution on $J_k$ and $J_1$ is given by \begin{equation}
\mathrm{P}(J_k < y, J_1 < x) = \left(1-y^{-\alpha}\right)^N \sum_{j=0}^{k-1} \left(\begin{array}{c} N \\ j \end{array}\right) \left(\frac{y^{-\alpha} - x^{-\alpha}}{1-y^{-\alpha}}\right)^j
\end{equation}To very good approximation when $N$ is large, we have \begin{align}
\mathrm{P}\left(\frac{J_k}{J_1} > z\right) &= \int\limits_1^\infty \mathrm{d}y \left.\frac{\partial }{\partial y} \mathrm{P}(J_k<y, J_1<x)\right|_{x=y/z} \notag \\
&\approx \int\limits_1^\infty \mathrm{d}y \frac{\nu}{y^{\nu +1}} \mathrm{e}^{-Ny^{-\nu}} \left[N \sum_{j=0}^{k-1} \frac{N^j}{j!} \left(y^{-\nu} - y^{-\nu}z^\nu \right)^j   - \sum_{j=1}^{k-1} \frac{N^j}{(j-1)!} \left(y^{-\nu} - y^{-\nu}z^\nu \right)^{j-1} \right]\notag \\
&\approx \int\limits_1^\infty \mathrm{d}y \frac{\alpha N}{y^{\nu+1}} \mathrm{e}^{-Ny^{-\nu}} \frac{N^{k-1}}{(k-1)!} y^{-\alpha(k-1)} \left(1-z^\nu\right)^{k-1} = \int\limits_0^N \mathrm{d}u \mathrm{e}^{-u} \frac{u^k}{(k-1)!} \left(1-z^\nu\right)^{k-1} \notag \\ 
&\approx \left(1-z^\nu\right)^{k-1}   \label{jkoverj1}
\end{align}From here we can compute \begin{equation}
\left\langle \frac{J_k}{J_1}\right\rangle = \int\limits_0^1 \mathrm{d}\left(1-z^\nu\right)^{k-1} z  =  \frac{\Gamma(1+1/\nu)\Gamma(k)}{\Gamma(k+1/\nu)}.
\end{equation}This asymptotes to \begin{equation}
\left\langle \frac{J_k}{J_1}\right\rangle  \sim \frac{1}{k^{1/\nu}}.
\end{equation}
\subsection{The Thermodynamic Limit}
Now, let us study what happens in the large $N$ limit.  In particular, we want to understand the distribution on $K_1$.   As we will confirm in the next subsections, whenever $K_1 \gg 1/N$ the mean field description of the binary decision model breaks down.   Therefore, in this subsection, we will determine $\mathrm{P}(K_1 >0)$ (after the $N\rightarrow \infty$ limit).

Let's begin with the case $0<\nu <1$.   In this case, we can prove that $\mathrm{P}(K_1>0)=1$ by showing that $\langle 1/K_1\rangle$ is finite in the limit $N\rightarrow \infty$: \begin{equation}
\left\langle \frac{1}{K_1}\right\rangle = \sum_{k=1}^N \left\langle \frac{J_k}{J_1} \right\rangle \sim \sum_{k=1}^N \frac{1}{k^{1/\nu}} < \infty
\end{equation}  We thus see that almost surely there is no mean field limit in this case.

Next, let us look at the case $\nu=1$.   In this case, we see that $J_1$ is almost surely of order $N$.\footnote{Rigorously, we mean that $\mathrm{P}(J_1/N > r)$ goes to 0 for $r\rightarrow \infty$, and 1 as $r\rightarrow 0$, after we have taken the limit $N\rightarrow \infty$.}    The sum of all of the smaller random variables is of the order \begin{equation}
\left\langle \sum_{i=2}^N J_i \Theta(J_1-J_i)\right\rangle \sim N\int\limits_0^{J_1} \frac{\mathrm{d}J_1}{J_1} \sim N\log J_1 \sim N\log N
\end{equation}with standard deviation \begin{equation}
\left[\mathrm{Var}\left(\sum_{i=2}^N J_i \Theta(J_1-J_i) \right)\right]^{1/2}\sim (NJ_1)^{1/2} \sim N.
\end{equation}Thus in the strict limit $N\rightarrow \infty$ we see that there is a mean field limit almost surely.    However, this is a very slow limiting process, and we will return to this point shortly.

The case $\nu>1$ also has a mean field limit, as it should.   To explicitly check this, we see that $N^{1/\nu} \sim J_1 \ll J_{\mathrm{tot}}\sim N$, and \begin{equation}
\mathrm{Var}(J_{\mathrm{tot}})^{1/2} \sim N^{1/2} \left[\int\limits_1^N \mathrm{d}J_1 J_1^{1-\nu}\right]^{1/2} \sim \left\lbrace\begin{array}{ll} N^{1/\nu} &\ 1<\nu <2 \\ \sqrt{N\log N} &\ \nu = 2 \\ \sqrt{N} &\ \nu >2 \end{array}\right..
\end{equation}When $\nu>2$, of  course, the standard central limit theorem holds and $J_{\mathrm{tot}}$ is essentially a Gaussian distribution.   Regardless, we see that for $\nu>1$, fluctuations are subleading just as we expect, so the mean field limit exists.

We should also note that even if there is formally a mean field limit, this mean field limit may require $N$ to be so large that it is impractical for ``realistic" situations.    To that end, let us alternatively say that the mean field limit occurs when there are almost surely (away from any possible phase transitions) no avalanches of size larger than $\lambda$.    In any linear response regime, we have shocks no larger than $\Delta q \sim K_1/(1-\alpha)$, which simply follows from the fact that having many influential nodes all have very similar $P$ values is very unlikely.   Note that in the case where there is no influence, $K_1=1/N$, and the total number of nodes that flips on average reduces to $1/(1-\alpha)$, as was discussed in Section \ref{sec2}.   We conclude that when $K_1 \ll \lambda$, we have a mean field limit, and when $K_1 \gg \lambda$, we do not.\footnote{The $\alpha$ scaling has been dropped for convenience, since our main interest here is the scaling with system size $N$.}    

Now, we have already seen that $K_1$ is almost surely positive when $\nu<1$, so there can be no mean field limit here, so long as $\lambda$ is chosen to be small enough.     In the case $\nu=1$, we have $K_1 \sim 1/\log N$, which means that if we are looking at scale $\lambda$, the size of $N$ required to have a mean field limit is $N\sim \exp[1/\lambda]$, which is likely much larger than any practical graph for any reasonable resolution $\lambda$.    When $\nu >1$, we require $N^{1/\nu - 1} \ll \lambda$, or $N\gg (1/\lambda)^{\nu/(\nu-1)}$, which may be ``surprisingly" large.   For example, in the case $\nu=2$ which is believed to be experimentally relevant in many systems, we actually need to have $N\sim 1/\lambda^2$ to have a mean field limit, and if our resolution $\lambda$ is small, this limit may be beyond our reach with any practical social system.   We should finally note that the limit $\nu \rightarrow \infty$ results in $N\sim 1/\lambda$, which is trivial since the number of nodes which flip in any avalanche must be an integer multiple of $1/N$.     This final case where $N\sim 1/\lambda$ will hold, up to minor corrections, in most systems where $J_1$ does not have a power law tail distribution:   for example, in the case of a Gaussian distribution on $J$s, we have $K_1 \sim N^{-1}\sqrt{\log N}$, which is not much different from $N^{-1}$.

\section{The Complete Graph}\label{sec4}
In this section, we begin by studying the binary decision model with very heterogeneous influence ($\nu<1$) on the complete graph.   The result will be an unpredictable phase characterized by abrupt shocks without critical phenomena, and with no thermodynamic limit with large $N$.   This is a ``solvable" model in the sense that the dynamical behavior may be completely explained, and certain quantities can, in certain limits, be computed exactly, as we describe in this section.
\subsection{A Picture Gallery}
\begin{figure}[here]
\centering

   \includegraphics{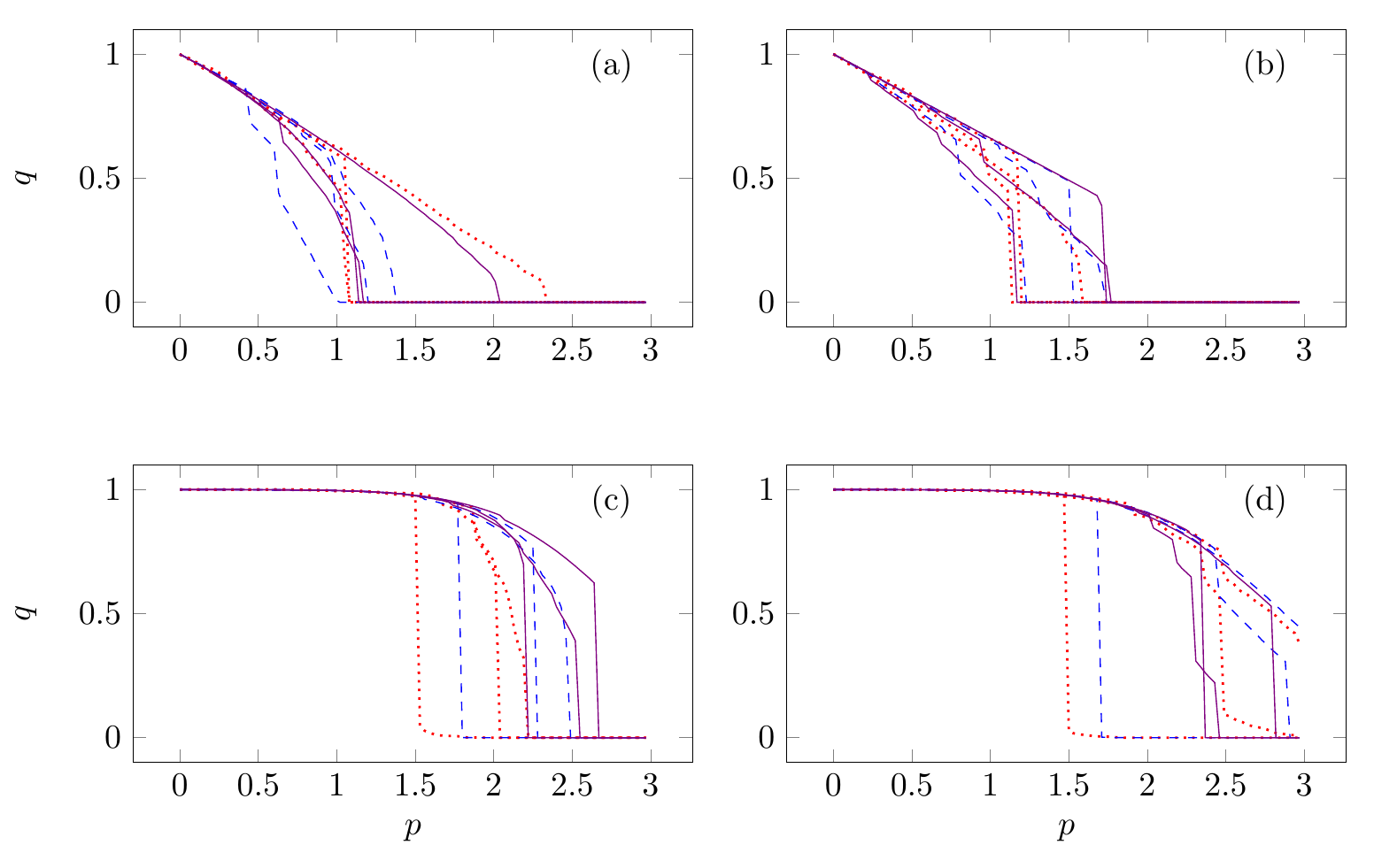} 
   
\caption{We show typical $q(p)$ for unpredictable decision making on complete graphs.   Plots (a) and (c) have $\nu=0.9$, and (b) and (d) have $\nu=0.5$.   Plots (a) and (b) have $P_v$ uniformly distributed on $[0,1]$, and (c) and (d) have $P_v$ normally distributed with mean 1 and standard deviation 0.25.    Dotted lines correspond to graphs with $N=10^3$; dashed lines to $N=10^5$; solid lines to $N=10^7$.   To avoid clutter, we did not display the lower branch of the bistable decision making.}
\label{fig1}
\end{figure}

As we demonstrated in the previous section, there can be no mean field limit in the case where $\nu<1$.    The consequences of the mean field limit being non-existent are dramatic:  the macroscopic  dynamics of the system are sensitive to the realization of $P_v$, which is microscopic data.   This should make sense, since when $\nu<1$ a couple nodes can carry quite a large fraction of the influence in the graph.   So we begin by showing in Figure \ref{fig1} some typical trajectories of $q(p)$, found by increasing $p$ starting from $q(0)=1$.   Since we are working on a complete graph, we can write very simple code by using the fact that we can order the nodes in order of their values of $P_v$, and sequentially look through them to determine whether or not they will be in state 0 or 1 at a given value of $p$.   Exploiting the ``graph structure" here allows us to work with orders of magnitude more nodes with fast run times.   Note that despite $N$ ranging over 4 orders of magnitude, there is no noticeable mean field limit, just as we predicted earlier.   Also note, particularly on the uniform distribution, the possibility of multiple large shocks.

There have been previous models which admit shocks without critical phenomena -- admitting lower cutoffs in the distribution of $P_v$ can result in sudden shocks, as was demonstrated in \cite{Lucas2013, Pradhan2005}.    However, the key difference with these models is that in the former models, the location of this shock is predictable given the microscopic dynamics.   Furthermore, the shock in these models is typically of the form $q(p) = \Theta(p_{\mathrm{c}}-p)$, which means that the entire market  changes its state in one global avalanche.    Although this possibility should not be ruled out, since some economic markets (e.g., housing) will never undergo such radical binary shifts, alternative mechanisms for unpredictability should be explored.    One mechanism which is slightly simpler mathematically than heavy-tailed influence distributions is to allow the function $F(P)$ distribution to have discontinuous first derivative (or even be discontinuous) at points where $0<F(P)<1$.   However,  this mechanism seems rather non-robust and would require the model to be fine-tuned.
\subsection{Analytic Results}
Now, let us return to unpredictable decision making, and focus on some analytic computations which help illuminate the underlying behavior of these systems.   The picture gallery makes clear that the various phases of this model, encapsulated in the double-valued $q(p)$ is itself a random variable in unpredictable decision making.    However, this does not imply that the model cannot be solved in any limit;   instead, what we mean by ``solvability" must be slightly re-defined.   This is precisely analogous to the way in which the fair coin flip problem is exactly solved -- there is an equal probability for the coin to give heads or tails, although on a given realization the result will be unknown.     Similarly, for this problem, the  easily solvable limit corresponds to $\nu \rightarrow 0$, in which only one node should dominate the overall influence, and therefore the macroscopic behavior.    Higher order corrections will become more and more relevant for larger $\nu$, in which the behavior of two, three, etc. nodes must be accounted for.

In this limit, it is very easy to understand the model, to first order.  The first order approximation for this computation is that exactly one node contributes to $Q$:   i.e. \begin{equation}
Q = x_1.
\end{equation}Here $x_1$ refers to the state of the most influential node.  The decision making will be bistable and will correspond to solutions of the pair of equations \begin{subequations}\begin{align}
x_1 &= \Theta(P_1 h(x_1) - p), \\
q &= F\left(\frac{p}{h(x_1)}\right).
\end{align}\end{subequations}From these equations it is straightforward to derive that $q(p)$ is a multivalued function with upper branch $q_+(p)$ and lower branch $q_-(p)$: \begin{subequations}\begin{align}
q_+(p) &= \left\lbrace \begin{array}{ll} F(p/h(1)) &\ P_1 h(1) < p \\ F(p) &\ P_1 h(1)> p \end{array}\right., \\
q_-(p) &= \left\lbrace \begin{array}{ll} F(p/h(1)) &\ P_1  < p \\ F(p) &\ P_1 > p \end{array} \label{multivalq}\right.,
\end{align}\end{subequations}insofar as the first order approximation that only one node's influence is relevant.   Indeed, we see that the location of the shocks between the two branches of $q(p)$ is completely unpredictable, as is the location of the shock.   However, given the location of one shock, the location of the other is determined, as are the sizes of both shocks -- everything is determined by the random variable $P_1$  -- the intrinsic value (internal magnetic field) as determined by the most influential node in the graph.

 In this section, our main numerical result will be to compute the probability density function of $s_1$, which is the size of the shock during which the most influential node flips from 1 to 0.   The size of the shock is easily shown to be \begin{equation}
s_1 = F(P_1) - F(h(1)P_1) \label{s1firstorder}
\end{equation}and therefore the distribution on $s_1$ can be found straightforwardly by studying the preimage of the function of the right hand side.

In general it is quite ugly to invert Eq. (\ref{s1firstorder}).   In the case where $P_1$ is distributed uniformly on $[0,1]$ and $h=1+AQ$, however, it is straightforward to compute \begin{equation}
s_1 = \left\lbrace\begin{array}{ll} AP_1 &\  P_1 < 1/(A+1) \\ 1-P_1 &\ P_1 \ge 1/(A+1) \end{array}\right..
\end{equation}This implies that $s_1$ is uniformly distributed on $[0, A/(A+1)]$.

\begin{figure}[here]
\centering

   \includegraphics{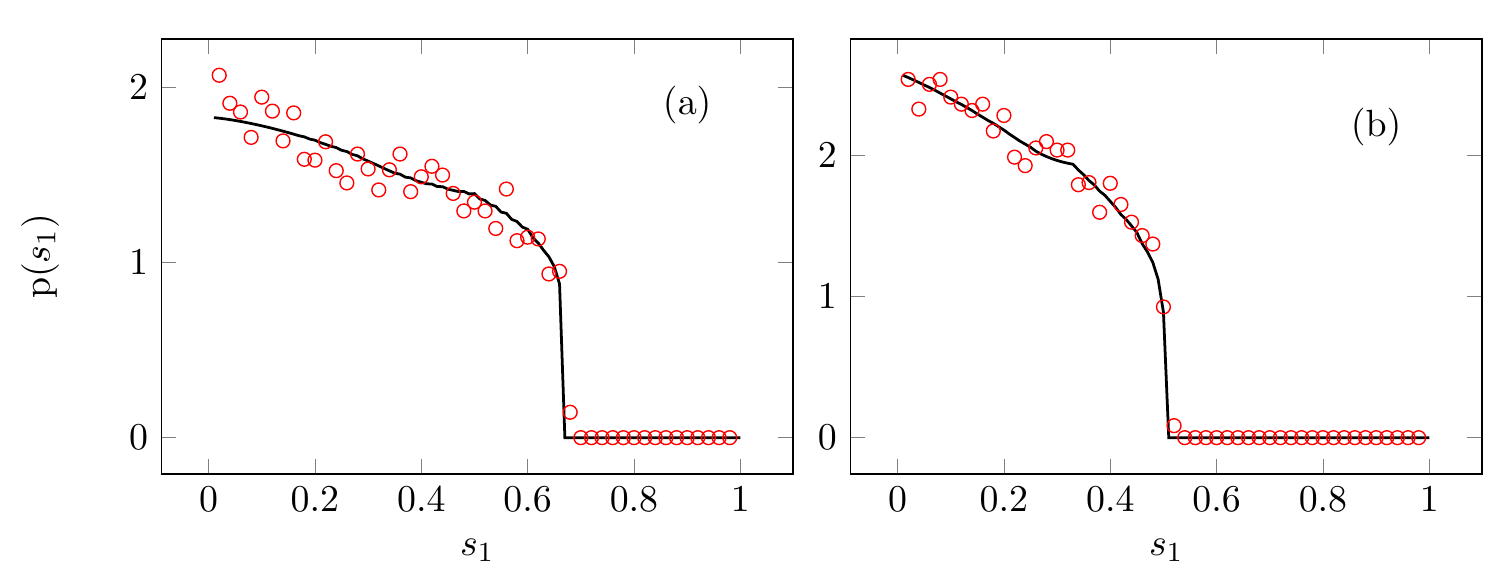} 
   
\caption{We show the theoretical estimate for the PDF $\mathrm{p}(s_1)$ (solid line) given by Eq. (\ref{eq26})  and the numerical estimate (circles) for the specific case $\nu=0.3$.  The numerical estimate was obtained from $10^4$ trials, binning data points into 50 bins, on graphs with $N=50000$ nodes.   (a) shows the case $A=2$;  (b) shows the case $A=1$.}
\label{fig2}
\end{figure}

From the picture gallery, it is clear that in general a first order approximation is not enough.    This is because there are multiple jumps (shocks) in many of the curves (note that the number of large jumps is random as well), and whenever there are two or more shocks, that implies that at least two influential nodes were relevant.\footnote{Note that there can be two influential nodes which contribute to the same shock, as well, so the number of shocks does not correspond to the number of ``very influential" nodes.} We will see that numerical data demonstrates this is true as well for the distribution of $s_1$.     Let us therefore also describe the second order approximation, which already becomes quite cumbersome.   We can tackle this most straightforwardly by first breaking the problem into two problems, each of which happens with half probability:   $P_1<P_2$ and $P_2 <P_1$.    Since $K_1=1-K_2$, $P_1$ and $P_2$ are each independent, we will assume $K_1$ is fixed and average over its distribution at the very end.    To begin, let us denote $P_- = \min(P_1,P_2)$ and $P_+ = \max(P_1,P_2)$.   We can write down, the probability density function on the shock size $s_1$ by computing the contributions from the probability that $P_1<P_2$ and both nodes are in different shocks, then the same shock;   and repeating for $P_2<P_1$: \begin{align}
\mathrm{p}(s | K_1) &=  \int\limits_{P_- < P_+} (-F^\prime(P_-) \mathrm{d}P_-)(-F^\prime(P_-) \mathrm{d}P_+)\left[ \delta\left(s-F(P_-)+F\left(\frac{h(1)}{h(K_2)}P_-\right)\right) \Theta\left(P_+ - \frac{h(1)}{h(K_2)} P_-\right) \right. \notag \\
&\left. + \delta\left(s-F(P_-)+F\left(h(1)P_-\right)\right) \left( \Theta\left( \frac{h(1)}{h(K_2)} P_--P_+\right) +  \Theta\left( \frac{h(1)}{h(K_1)} P_--P_+\right)   \right)\right.  \notag \\
&\left. + \delta\left(s - F\left(\frac{h(1)}{h(K_1)} P_+\right) + F(h(1)P_+)\right) \Theta\left(P_+ -  \frac{h(1)}{h(K_1)} P_-\right)    \right].
\end{align} In the case of $P_1$ and $P_2$ uniformly distributed on $[0,1]$ with $h(Q) = 1+AQ$, this integral can be explicitly evaluated: \begin{align}
\mathrm{p}(s | K_1) &= \Theta(s)\Theta\left(\frac{AK_1}{1+A}-s\right)\left(\frac{1+A(1-K_1)}{AK_1} + s \left(1 - \frac{(1+A)(1+A(1-K_1))}{(AK_1)^2}\right) \right) \notag \\
&+ \Theta(s) \Theta\left(\frac{A}{A+1}-s\right)\left(\frac{1-K_1}{1+AK_1} + \frac{K_1}{1+A(1-K_1)}\right)\frac{s}{A}  \notag \\
&+ \Theta\left(\frac{A}{A+1}-s\right)\Theta\left(s-\frac{AK_1}{1+A}\right) \frac{AK_1(1-s)}{1+A(1-K_1)} + \Theta(s)\Theta\left(\frac{A(1-K_1)}{1+A}-s\right)s \notag \\
&+ \Theta\left(s-\frac{A(1-K_1)}{1+A}\right)\Theta\left(\frac{A}{A+1}-s\right) \frac{A(1-K_1)(1-s)}{1+AK_1} \notag \\
&+ \Theta(s)\Theta\left(\frac{AK_1}{1+AK_1}-s\right) \left(\frac{1+AK_1}{1+A}\right)^3 \left(1+ s\left(\frac{1}{(AK_1)^2}-1\right)\right). \label{eq26}
\end{align}At this point, we simply perform the average over $K_1$ numerically, by using Eq. (\ref{jkoverj1}).    Figure \ref{fig2} shows that for a small value of $\nu$, there is excellent agreement between the theoretical prediction and numerical data, suggesting that in this limit, only the first two nodes are relevant.   Note that the first order approximation would \emph{not} have sufficed.

\section{Random Graphs}\label{sec5}
Now let us turn to the behavior of this model on random graphs.    In \cite{Lucas2013} it was shown that in the case of no influence, the presence of a network turned the mean field model into a glassy model with exponential numbers of equilibria.   In this section, we will see that models which are unpredictable on complete graphs become predictable again on random graphs with finite degree, but enter a new glassy phase distinct from the mean field glassy phases.
\subsection{Glassy Behavior}
Glassy behavior comes about from the possibility that a cluster of nodes could simultaneously be all in state 1 or all in state 0.   This is especially elegant on locally tree-like graphs, where there is a very simple interpretation of the typical cluster as a pair of spins which simultaneously flip, followed by the resulting avalanche \cite{Lucas2013}.   However, this picture breaks down in the very heterogeneous influence model, because most nodes are only following a single node.   This must remain true on random graphs, because the L\'evy flights associated to the total influence that each node sees are dominated by a few terms regardless of the number of terms in the sum. 

Let us assume that the random graph is sparse in that the degree of any node is much smaller than the total number of nodes.   Let us also assume that $\nu$ is close to 0 -- we will return to what happens at larger values of $\nu$ later.   Although the  most influential node cannot influence the entire graph as it can on the complete graph, it does still entirely determine the state of each of its neighbors.    Now, let us imagine removing the influential node and all of its neihgbors from the graph.   We can perform the same procedure again, and remove more nodes which are mostly entirely following a different node.   Continuing this process until the graph has no remaining nodes, we see that the graph decomposes into local clusters which are effectively described by the unpredictable model on a complete graph.   Since we have many of these clusters, each of which have independent thresholds, however, we expect a well-behaved and predictable thermodynamic limit to emerge.

Let us be more quantitative about this resulting limit.   To do this, let us first focus on a single cluster.   Within a single cluster, the state is entirely determined by $P_1$, the internal field of the most influential node.    Each cluster is characterized by a configuration with a regime of bistability at intermediate $p$, which can be described by specifying the upper and lower branches of the multivalued function $q(p)$, which is precisely the result we found earlier in Eq. (\ref{multivalq}).   Now, let us average over the value of $P_1$.    This average is actually quite simple to take: \begin{subequations} \label{gmf}\begin{align}
q_+(p) &= F\left(\frac{p}{h(1)}\right)^2 + \left[1 - F\left(\frac{p}{h(1)}\right)\right] F(p), \\
q_-(p) &= F(p) F\left(\frac{p}{h(1)}\right) + \left[1 - F(p)\right] F(p).
\end{align}\end{subequations}
So we conclude that if $F(p)$ is a continuous function, this model is in fact characterized by a \emph{single, glassy phase} with equilibrium spectrum width \begin{equation}
\Delta q(p) \equiv q_+(p) - q_-(p) = \left[F\left(\frac{p}{h(1)}\right) - F(p)\right]^2.
\end{equation}We emphasize that this glassy phase not only is independent of $N$, as it must to be a well-behaved thermodynamic limit, but it is entirely independent of the graph, so long as the degree of each node is much smaller than $N$.

\begin{figure}[here]
\centering

   \includegraphics{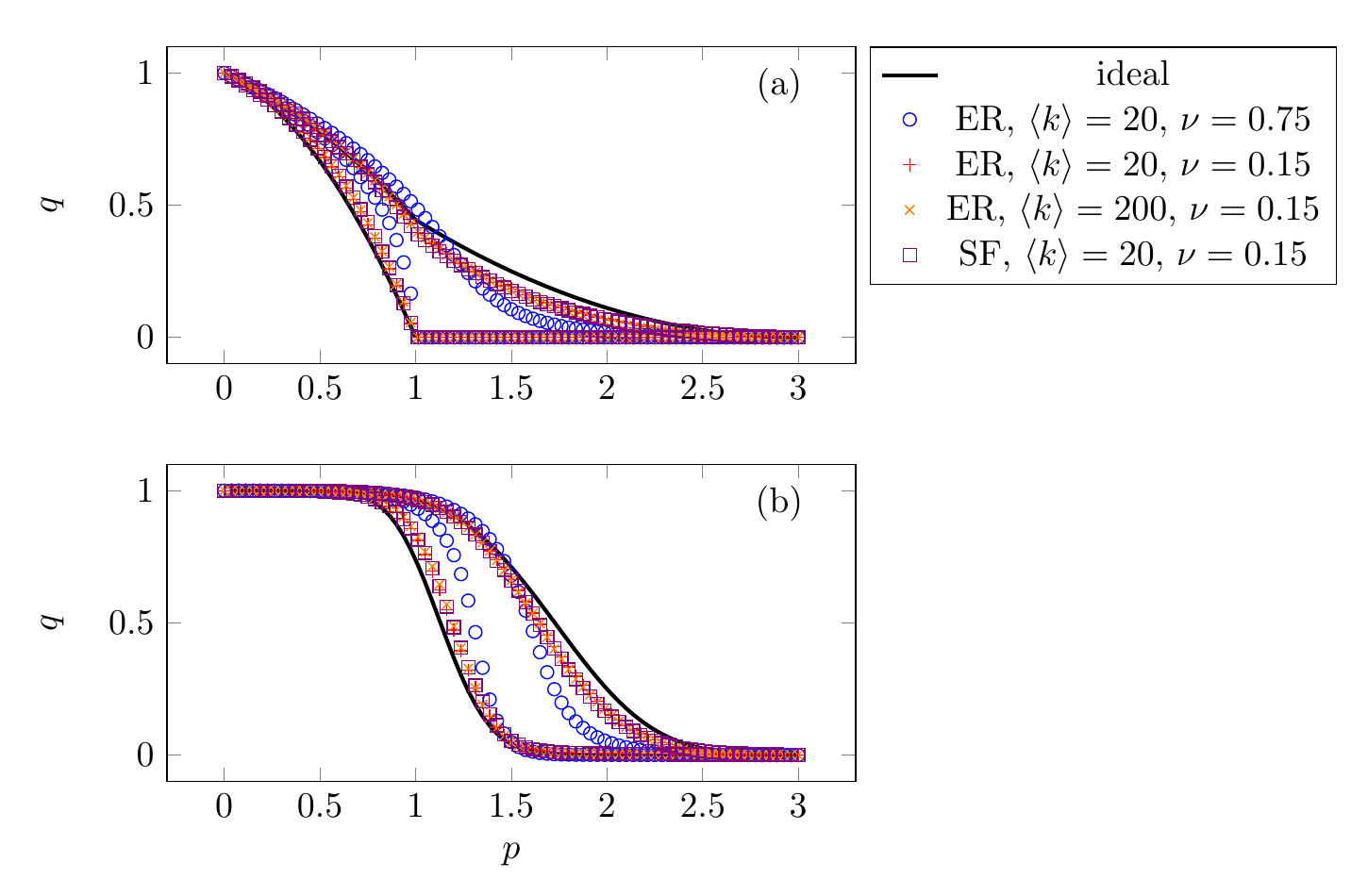} 
   
\caption{We compare the theoretical glassy phase to the numerical one on various graphs and values of $\nu$.   In the legend, ER stands for Erd\"os-R\'enyi graphs, and SF stands for scale free graphs with degree distribution $\rho_k \sim k^{-3}$.  Plots (a) has a uniform distribution for $P_v$ and $A=2$, and plot (b) has a Gaussian distribution with $\mu=1$, $\sigma=0.25$, and $A=1$.   All graphs have $N=2000$ nodes.   We averaged over 200 trials for each data point.}
\label{avgruns}
\end{figure}

\begin{figure}[here]
\centering

   \includegraphics{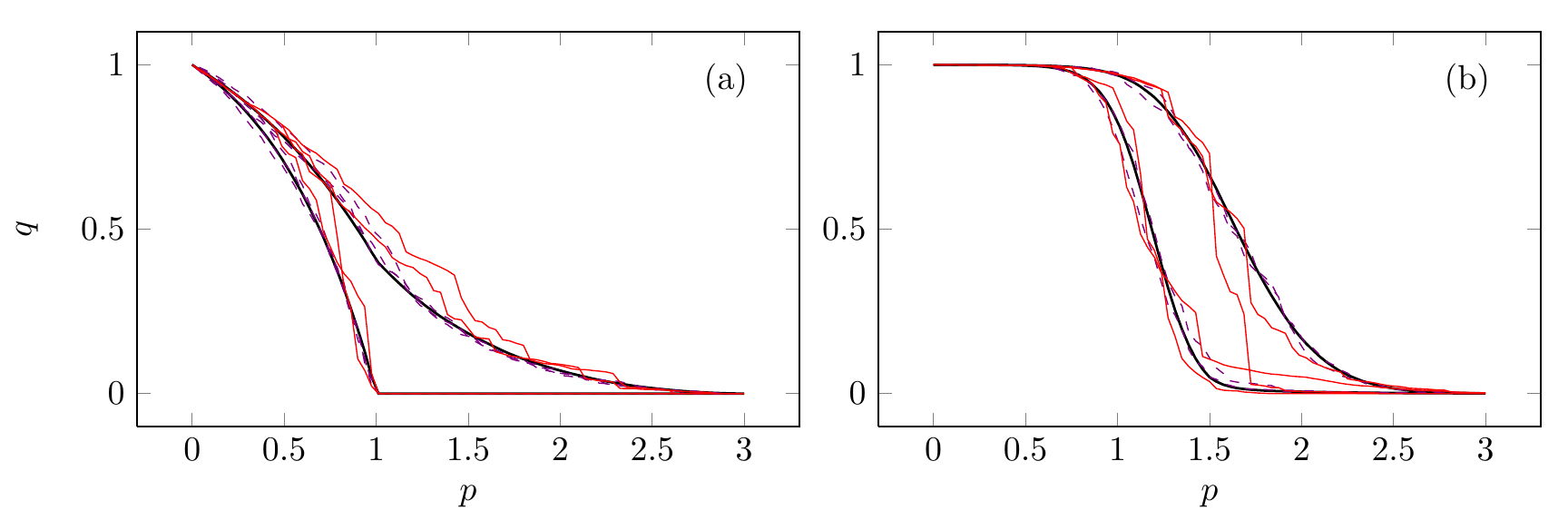} 
   
\caption{We compare samples of influential decision making on graphs to the average behavior.   We use Erd\"os-R\'enyi graphs with $N=2000$ nodes, and influence distributions with $\nu=0.15$.    Plots (a) has a uniform distribution for $P_v$ and $A=2$, and plot (b) has a Gaussian distribution with $\mu=1$, $\sigma=0.25$, and $A=1$.   The thick solid line corresponds to the average case;  thin solid lines refer to graphs with $\langle k\rangle = 200$, and dashed lines to graphs with $\langle k\rangle=20$.}
\label{trialruns}
\end{figure}

Figure \ref{avgruns} demonstrates that qualitatively, the theory is very accurate at small values of $\nu$, and becomes less so at larger values of $\nu$.   We will return to this point further in a later section.   Note that the glassy phase described above is remarkably independent of the underlying network structure, just as we predicted.   Figure \ref{trialruns} shows typical samples of the total glassy phase at finite values of $\langle k\rangle$, when compared to the approximately $\langle k\rangle$-independent result.    

\subsection{Finite $N$ Fluctuations}
\begin{figure}[here]
\centering

   \includegraphics{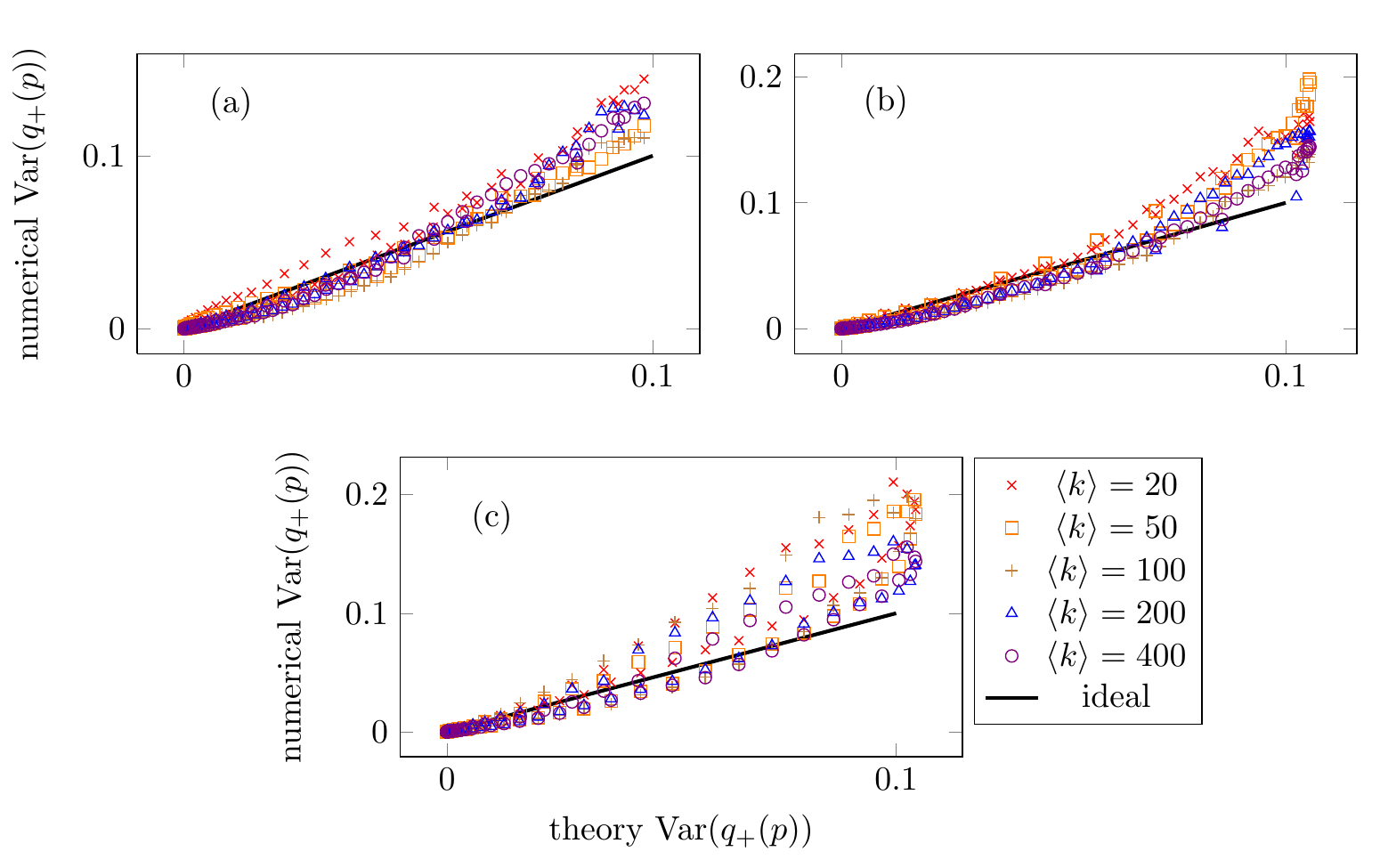} 
   
\caption{We compare the prediction of Eq. (\ref{eq36}) to numerical results on Erd\"os-R\'enyi graphs with $N=2000$ nodes, and influence distributions with $\nu=0.15$.    Plots (a) and (b) have uniform distributions for $P_v$, and plot (c) has a Gaussian distribution with $\mu=1$, $\sigma=0.25$.   Plot (a) has $A=2$, (b) has $A=4$, (c) has $A=1$.   In each case, we have multiplied the numerical variance by $N/\langle k\rangle$ to stress that after correcting for this graph dependent pre factor, the qualitative results are independent of graph.    Variances were obtained by averaging over 200 trajectories $q_+(p)$ for each instance.  }
\label{varfigure}
\end{figure}

At a finite value of $N$, there will of course be fluctuations in mean field theory.   For each cluster, the variance is given by \begin{subequations}\begin{align}
\mathrm{Var}(q_+(p))_1 &= F\left(\frac{p}{h(1)}\right) F\left(\frac{p}{h(1)}\right)^2 + \left[1- F\left(\frac{p}{h(1)}\right) \right]F(p)^2 - \left[F\left(\frac{p}{h(1)}\right)^2 +\left[1- F\left(\frac{p}{h(1)}\right) \right] F(p)\right]^2   \notag \\
&= F\left(\frac{p}{h(1)}\right) \left[1-F\left(\frac{p}{h(1)}\right)\right]\left[F\left(\frac{p}{h(1)}\right)-F(p)\right]^2  \\ 
\mathrm{Var}(q_-(p))_1 &= F\left(p\right) F\left(\frac{p}{h(1)}\right)^2 + \left[1- F\left(p\right) \right]F(p)^2 - \left[F\left(p\right)F\left(\frac{p}{h(1)}\right) +\left[1- F\left(p\right) \right] F(p)\right]^2 \notag \\
&= F(p)(1-F(p)) \left[F\left(\frac{p}{h(1)}\right)-F(p)\right]^2 .
\end{align}\end{subequations}
For the total graph, we just need to divide by the number of clusters, which is of order $N/\langle k\rangle$, as each cluster behaves independently, so we expect \begin{equation}
\mathrm{Var}(q(p)) \sim \frac{\langle k\rangle}{N} \mathrm{Var}(q(p))_1.\label{eq36}
\end{equation}Surprisingly, for many regions of parameter space this relation is obeyed fairly quantitatively (to within a factor of 2), despite our crude estimate of the number of clusters, as shown in Figure \ref{varfigure}.  The scaling in $\langle k\rangle$ is always obeyed although sometimes the effective prefactor appears to differ -- depending on what the internal distribution $F(P)$ is.

\subsection{Phase Transitions Suppressed by Influence}
Let us now compare the glassy phase of the very heterogeneous influence model with the other case -- where influence is not ``heavy-tailed".   Making precise what we mean by ``heavy-tailed" is likely quite cumbersome, and for this paper we will simply continue to refer to $J$ distributions with $\nu<1$ as heavy-tailed.

To understand what happens when the influence distribution is not heavy-tailed, consider the opposite limiting case where $J\approx 1$, with small amounts of disorder in influence.    Influence in this limit  has \emph{almost no effect} on the equilibria.   We have seen this already at mean field level, but it also holds to higher orders in the cavity expansion.   We say ``almost" here because there is a very small effect, which is analogous to the effect of small degrees in the homogeneous influence model.       To write the cavity equation, let us denote the state of the cavity node with $x_0$, and one of its neighbors with $x_1$.   Let $K_{10}$ denote the relative influence of the cavity on its neighbor.   Averaging over the neighbors of the cavity, we find analogously to the computation in \cite{Lucas2013} that \begin{equation}
\mathrm{P}(x_1=1) = q + \alpha \langle K_{10}|J_0\rangle (x_0-q),
\end{equation}and from the $x_0$-dependent piece of this equation, we can extract that \begin{equation}
\mathrm{P}(x_0=0,1\text{ are both possible}) = \alpha^2 \langle K_{10}\rangle,
\end{equation}where we have averaged over $P_0$ and $J_0$, about the mean field solution.   But $\langle K_{10}\rangle = 1/\langle k\rangle$, a fact which is easy to see by the following logic: \begin{equation}
\langle \langle K_{10}|J_0\rangle_{k_1} \rangle_{J_0} = \langle \langle K_{10} | k_1\rangle_{J_0} \rangle_{k_1} = \langle k_1^{-1}\rangle_{k_1} = \frac{1}{\langle k\rangle}.
\end{equation}We have used the tower property of conditional expectation values here, and used that averaging over all influences, the influence of any one neighbor is equally likely to that of any other, and therefore the influence must be the inverse of the degree.    We conclude, using this logic, that the first order cavity correction to the mean field equations is therefore independent of influence.   Of course, we showed with a heuristic argument earlier that very heterogeneous influence models do alter the glassy physics, and confirmed this numerically.   The break down in this argument is simply that we cannot linearize around a solution in the very heterogeneous influence model, since the behavior of $x_0$ is entirely dominated by the most influential node.

\begin{figure}[here]
\centering

   \includegraphics{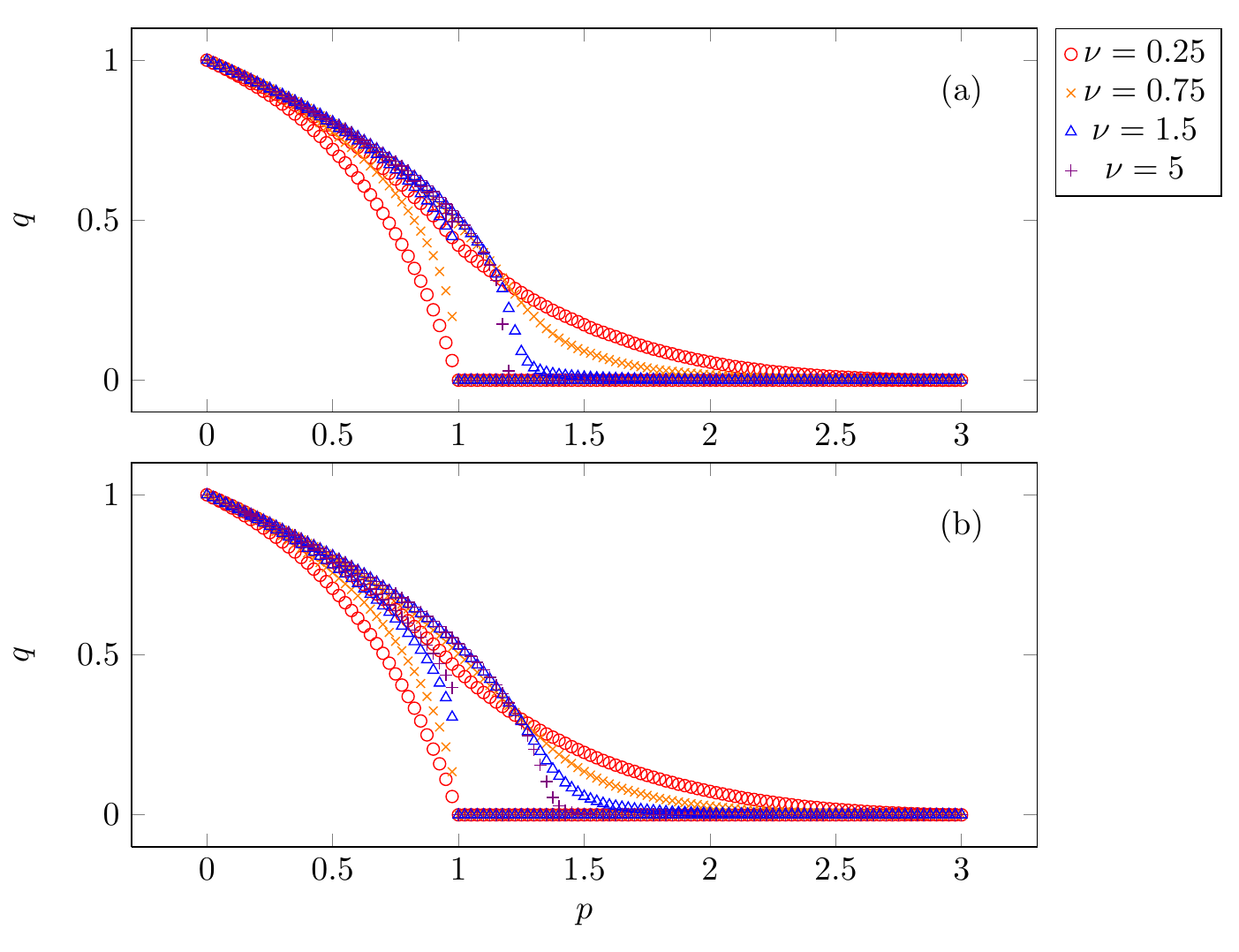} 
   
\caption{We show the maximal and minimal values of $q(p)$ for various values of $\nu$ and $\langle k\rangle$.   (a) shows Erd\"os-R\'enyi graphs with $\langle k\rangle =40$, and (b) shows $\langle k\rangle = 10$.   All runs had uniform distributions on $P_v$ with $A=2$ and $N=2000$ nodes.  100 trials were averaged to obtain each data point.   We have also checked that these results are $N$-independent.}
\label{fig6}
\end{figure}

So we conclude that for $\nu \ll 1$, we should have the highly glassy phase described in earlier subsections, and for $\nu \gg 1$ we should return to the glassy phase of \cite{Lucas2013}, which has small $1/\langle k\rangle$ deviations from mean field theory (with the possible suppression of discontinuities in phase transitions).   Figure \ref{fig6} shows the transition between the two glassy phases, as $\nu$ is varied, for two values of $\langle k\rangle$.\footnote{Due to the fact that we have allowed nodes to count their own opinion in $Q_v$, the spectrum $\Delta q$ of the glassy phase in the $\nu \gg 1$ limit picks up a factor of 4, when compared with the result of \cite{Lucas2013}:  $\Delta q \approx 4\alpha^2/(1-\alpha)\langle k\rangle$.   This comes about from the fact that each node in a glassy pair changes $Q_v$ by $2\alpha/\langle k\rangle$.}   It is likely that for $\nu \sim 1$ there is a complicated interplay between finite degree effects in the network and multiple influential nodes being relevant, and so we do not attempt a further theoretical analysis of this case.    We do make a final note, however:  we see numerical evidence that influence may suppress a phase transition even when $\nu>1$.    In the simulations with $\langle k\rangle =40$, there is a genuine phase transition at $\nu=5$; however, at $\nu=1.5$, we see a ``tail" in $q(p)$.   It is reasonable to suspect that a wide enough distribution in influence may thus have a benign effect on the global decision making process by helping to smooth out shocks and other unfortunate behavior.

\subsection{Correlated Degree and Influence}
In many cases, such as models of the type presented in \cite{Zhou2011}, it does not make sense to ignore correlations between degree and influence.   In this section, we will consider what happens when we consider binary decision making with influence which translates into graph structure according to some relation $k_v = k(J_v)$ with $k(J)$ a monotonically increasing function.    For this paper, we will focus on the simple function \begin{equation}
k(J) = k_0 + J^\theta  \label{kofj}
\end{equation} with $J$ distributed as usual according to Eq. (\ref{jdist}) and $0<\theta$ some power.    We will choose to construct our graphs as random graphs made with the Molloy-Reed algorithm \cite{Molloy1995, Newman2003}.  For large $k$, the degree distribution of the network will be given by \begin{equation}
\rho_k \sim k^{-1-\nu/\theta}.
\end{equation}Depending on the values of $\nu$ and $\theta$, we thus expect a variety of possible behaviors to occur.

If $\nu \le \theta$, we see that the distribution on $\rho_k$ is heavy tailed enough that some nodes in the graph should touch a finite fraction of the nodes.   This implies that very influential nodes have many connections.   if $\nu<1$, we also conclude that there is no mean field limit and therefore the decision making is unpredictable.    If $\nu>1$, then unless $\langle k\rangle$ is very small, we expect that we quickly transition into a homogeneous-like glass phase.  To see this, we note that \emph{all} of the most influential nodes are connected to each node, and since for $\nu>1$ the very influential nodes will count roughly equally, the independence of $P_v$ from $J_v$ implies that each node will see a crude ``mean field" collection of nodes.   If $\nu>\theta$, then the distribution on $\rho_k$ is not heavy tailed enough, and the most connected node will most likely not reach a finite fraction of nodes in the graph.   This implies that there is a mean field limit.    If $\nu<1$, the mean field limit will be qualitatively similar to the universal glassy phase described in the previous subsection, and if $\nu>1$, the mean field limit will be the homogeneous influence limit.   

\begin{figure}[here]
\centering

   \includegraphics{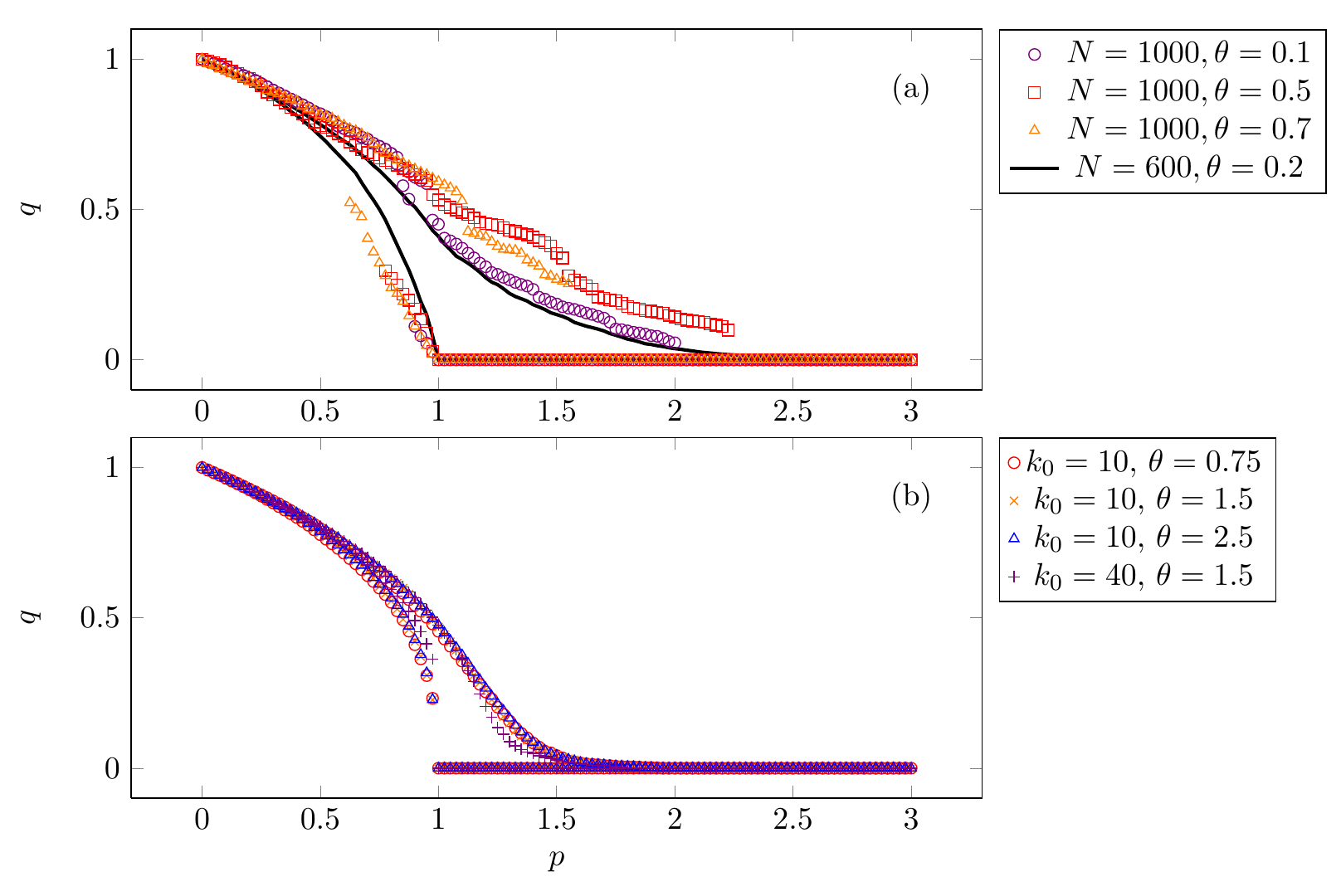} 
   
\caption{We show the maximal and minimal values of $q(p)$ for graphs where influence is correlated to degree.  (a) shows scale free graphs with $\nu=0.4$ and $k_0=10$; the solid line shows an average curve, and the marked lines show single instances of the model (recall that the average $q$ is identical for both phases in this limit, something which is true numerically as well).   (b) shows $\nu=1.5$ -- here, we averaged over runs, and each run is on graphs with $N=2000$.   All runs had uniform distributions on $P_v$ with $A=2$;  200 trials were averaged to obtain each data point, with the exception of the solo runs in (a).   We have also checked that these results are $N$-independent.   We collected more data than is displayed -- for clarity we have only displayed a small subset, but the qualitative trends hold for all of the data.}
\label{fig7}
\end{figure}

Figure \ref{fig7} shows that these qualitative arguments hold in practice in our simulations.   As in the previous section, ``far" from limiting cases these arguments do not hold perfectly, but it is clear what the basic mechanisms at play are.  When $\nu<1$, the average $q(p)$ is qualitatively very similar to the Erd\"os-R\'enyi case described in the previous sections, with  minor quantitative differences due to $\theta>0$.   As $\theta$ becomes larger than $\nu$, the size of the shocks becomes increasingly large, although due to numerical limitations we could not observe a ``sharp" transition between behaviors at $\nu=\theta$.  The $\nu>1$ case depends very little on the value of $\theta$, and only on $\langle k\rangle$.   In general, we see substantial robustness of the model to precise details of the graph structure, at least for the random graph ensembles we have studied.

\section{Conclusion}
In this paper, we have discussed the consequences of adding heterogeneous influence to the binary decision model of \cite{Lucas2013}.   Interestingly, we found that only in the limit where a few nodes have finite fractions of the total influence does the behavior dramatically differ from the case where there is no influence at all.   Although we have found that less-heavy-tailed distributions can also qualitatively alter the behavior of the model near phase transitions, it is analogous to how network structure can suppress a phase transition when $\langle k\rangle$ is small.  

When a few nodes dominate the overall influence of the graph, two new phases emerge:  one with unpredictable shocks, and one with a universal and highly ``glassy" phase of non-interacting clusters.  Although we have not explicitly checked whether the observed phenomena hold in more general models such as random field Ising models, it seems to reasonable to postulate that these phenomena are fairly generic, at least at the qualitative level.  We also stress that the results we have found for very influential decision making are robust against changes to the probability distribution on $J$, so long as they preserve the heavy-tailed nature of the distribution.

Most importantly, we comment on the possible relevance of this work to experiments.   As we have seen, the presence of very influential decision makers ($\nu<1$) in a graph has dramatic effects.    If the model we have presented is a qualitatively good description of real decision making, then these dramatic effects have \emph{experimentally observable consequences}.   If the influential decision makers also are connected to a finite fraction of nodes in the graph, the resulting shocks do not have associated critical phenomena.   If an experimentalist has good resolution in $p$ in observations of $q(p)$, this would be a very easy effect to observe in practice.    If the influential decision makers are connected only to small clusters of the graph, then the dynamics effectively splits into non-interacting clusters, and the macroscopic system does not undergo any phase transitions -- thus, the observation of a phase transition or shock would easily invalidate this scenario.   Note that determining whether a model without phase transitions contains very influential individuals or not is still rather subtle, at least purely from $q(p)$-- instead, it is likely that microscopic data would be required.  If the avalanche distribution contains an anomalously large number of avalanches of order $\langle k\rangle$, that would suggest that large clusters are switching states simultaneously, which is a strong indication of influential decision making.  

There are a few possible drawbacks to looking for experimental evidence of the present model.   Firstly, this paper has assumed that $J_v$ and $P_v$ are uncorrelated for each node -- possible correlations may lead to new phenomena, such as influential elites crashing a market even when a large majority of typical nodes would be willing to participate in the absence of such influence.   Secondly, this paper has assumed that the decision rules are the same for all nodes, but this may be a bad assumption (do influential nodes interact more strongly, e.g. through a larger value of $A$?).   The assumption that all the neighbors of node $v$ perceive the same influence $J_v$ may be incorrect.   Further study of generalized models which relax these assumptions is worthwhile.

\section*{Acknowledgements}\addcontentsline{toc}{section}{Acknowledgements}
A.L.  is supported by the Purcell Fellowship at Harvard.     He would like to thank the anonymous referees for helpful suggestions.
\bibliographystyle{plain}
\addcontentsline{toc}{section}{References}
\bibliography{agingbib}

\end{document}